# Comprehensive overview and assessment of miRNA target prediction tools in *human* and *drosophila melanogaster*

Muniba Faiza[1], Khushnuma Tanveer[2#], Saman Fatihi[2], Yonghua Wang[1], Khalid Raza[2]*

[1]School of Food Science and Engineering, South China University of Technology, Guangzhou, China 510640

[2]Department of Computer Science, Jamia Millia Islamia, New Delhi, India 110025

*kraza@jmi.ac.in

November 04, 2017

## Abstract

MicroRNAs (miRNAs) are small non-coding RNAs that control gene expression at the post-transcriptional level through complementary base pairing with the target mRNA, leading to mRNA degradation and blocking translation process. Any dysfunctions of these small regulatory molecules have been linked with the development and progression of several diseases. Therefore, it is necessary to reliably predict potential miRNA targets. A large number of computational prediction tools have been developed which provide a faster way to find putative miRNA targets, but at the same time their results are often inconsistent. Hence, finding a reliable, functional miRNA target is still a challenging task. Also, each tool is equipped with different algorithms, and it is difficult for the biologists to know which tool is the best choice for their study. This paper briefly describes fundamental of miRNA target prediction algorithms, discuss frequently used prediction tools, and further, the performance of frequently used prediction tools have been assessed using experimentally validated high confident mature miRNAs and their targets for two organisms *Human and Drosophila Melanogaster*. Both *Drosophila Melanogaster* and *Human* supported miRNA target prediction tools have been evaluated separately to find out best performing tool for each of these two organisms. In the human dataset, TargetScan showed the best results amongst the other predictors followed by the miRmap and microT, whereas in the *D. Melanogaster* dataset, MicroT tool showed the best performance followed by the TargetScan in the comparison of other tools.

## 1. Introduction

Micro RNAs (miRNAs) are short endogenous RNAs nearly 22 nucleotides long originating from the non-coding RNAs (Bartel, 2004). miRNAs were first identified in *Caenorhabditis elegans* in the year 1993 using genetic methods (Lee et al., 1993). miRNAs are expressed from long transcripts produced in animals, plants, viruses, and single-celled eukaryotes (Liu et al., 2012). miRNAs have become the focus of many research because of their significant role in degradation of mRNA, post-translational inhibition through complimentary base pairing (He & Hannon, 2004), and ability to control many biological processes such as homeostasis (Liu et al., 2012). miRNA regulates the target mRNA to make adjustments to the forming corresponding protein, which dysregulates the functions of miRNA, thereby leading to several human diseases (Bing et al., 2012). Cancer is the most common disease caused by miRNAs and their differential expression leads to different types of cancer such as lung cancer (Yanaihara et al, 2006), prostate cancer (Porkka et al., 2007), and ovarian cancer (Yang et al., 2008). miRNAs have also been implicated for causing neurological disorders such as Alzheimer's disease (Hébert et al., 2009), Schizophrenia (Beveridge et al., 2010), and multiple sclerosis (Cox et a., 2010). A large amount of miRNA data has been generated in recent years due to the major efforts in identifying their targets, and inferring their functions which is difficult to explore and assess by using only biological methods. Therefore, the computational methods in biological research provide statistical approaches to assess their quality and accuracy.

In the last few years, several computational tools have been developed for the prediction of miRNA targets, but prediction results greatly vary among these tools due to differences in their algorithms and training features. Therefore, it is difficult for a scientist to choose the best miRNA target prediction tool. In this paper, we have evaluated the performance of 11 miRNA target prediction tools for human as well as *Drosophila melanogaster* datasets providing the comprehensive summary of the considered tools, their target prediction assessment based on various metrics, including accuracy, number of targets predicted, sensitivity, specificity, true positive rates, false positive rates, and so on.

Many approaches have been made in the past few years to assess and evaluate the performance of existing miRNA target prediction tools. Mendes et al. (2009) evaluated the miRNA gene finding methods and target identification, reporting some problems in the existing methods. Bartel (2009) discussed the features of available miRNA target prediction tools, highlighting reasons for the differences among their performances, including recognition of the target nucleotide opposite to the miRNA first nucleotide. According to Bartel (2009), TargetScan rewards an 'A' across from the position 1, whereas other algorithms with seed pairing feature rewards a Watson-Crick (WC) match across this position (Krek et al., 2005; Lewis et al., 2005; Stark et al., 2005; Gaidatzis et al., 2007). Ruby et al., (2007) identified many conserved miRNAs through large-scale sequencing, which were not predicted by the tools. Alexiou et al., (2009) tested eight miRNA target predictors on a small datasets of five and sixty one miRNAs and proposed that the targets predicted by more than one algorithm are better than the other targets. Fan and Kurgan, (2014) studied seven miRNA predictors and the TargetScan and miRMap showed the overall high quality. They proposed that the prediction of target sites is more difficult than predicting the target genes due to the lower predictive quality of the prediction tools at the duplex level. Srivastava et al., (2014) evaluated the performance of eleven miRNA target predictors on the plant datasets. Akhtar et al., (2015) assessed the accuracy of miRNA predictors and reported the prediction of large number of false positives as the major flaw; but their accuracy in the prediction of true targets is still questionable. This review is a comprehensive analysis of the performance of the existing miRNA target prediction tools.

## 2. Common features of miRNA target prediction tools

Computational methods are used to identify that how miRNAs specifically targets the mRNAs. Following are a few common features on which most of the miRNA target prediction tools are based (Sarah et al., 2014).

### 2.1 Seed match

The region of miRNA starting from 5'-end to the 3'-end consisting of first 2-8 nucleotides is called the seed sequence (Lewis et al., 2005). It is considered as Watson-Crick (WC) match between a miRNA and its target by most of the prediction tools. An alignment between the miRNA and its target lying within the WC matching without any gaps in between is considered as the perfect seed match. Different algorithms consider different types of seed matches. The most commonly considered seed matches are as follows (Lewis et al., 2003; Kreck et al., 2005; Brennecke et al., 2005):

a. 6-mer: a perfect seed matching for six nucleotides between the miRNA seed and the mRNA.

b. 7-mer-m8: a perfect seed match between 2-8 nucleotides of miRNA seed sequence.

c. 7mer-A1: a perfect seed match between 2-7 nucleotides of miRNA seed sequence in addition to an A across the miRNA first nucleotide.

d. 8-mer: a perfect seed match between nucleotides 2-8 of miRNA seed sequence in addition to an A across the miRNA first nucleotide.

### 2.2 Free energy

It is a Gibb's free energy which is used as a measure of stability of miRNA structure by many tools. When a miRNA binds to the target mRNA resulting to a stable structure, it is considered as the most likely target of that miRNA. The reactions with more negative delta-G are less reactive, and therefore, have more stability. The hybridization of miRNA with its target mRNA provide information about the high and low free energy regions and delta-G predicts the strength of bonding between the miRNA and its target mRNA (Yue et al., 2009).

### 2.3 Conservation

It is the occurrence of a same sequence across the species. This feature analyzes the regions such as the miRNA, 3'-UTR, 5'-UTR. It has been found that seed region is more conserved than the other regions (Lewis et al., 2003). A small portion of miRNA which interacts with the target mRNA has conserved pairing which compensates for the mismatched seed and known as '3'-Compensatory sites' (Friedman et al., 2009). Conservation analysis helps to predict whether a predicted miRNA target is functional or not.

### 2.4 Site accessibility

It is the measure of the ease of miRNA by which it may locate its target mRNA to hybridize with it. The miRNA first binds to a short accessible region of a mRNA and then their hybridization are marked by the unfolding of the mRNA secondary structure after the completion of binding of the miRNA. Hence, to find the most probable target of the miRNA, the amount of energy required to make a site accessible is evaluated.

There are a few other features which are used in most of the target prediction algorithms. GU Wobble seed match calculates the chances of a G pairing with a U instead of C (Doench et al., 2004). Position Contribution determines the position of a target sequence within the mRNA (Grimson et al., 2007). Seed pairing stability is the free energy change calculated for a predicted duplex (Garcia et al., 2011). Target-site abundance determines the number of sites occurring in the 3'-UTR (Garcia et al., 2011). Local AU content is the concentration of A and U nucleotides which flank in the corresponding seed region (Friedman et al., 2009; Betel et al., 2010). 3'-Compensatory pairing is the pairing region (12-17 nts) in which the base pairs match with miRNA nucleotides.

## 3. miRNA databases

Basically, there are few online miRNA databases which provide all the experimentally validated miRNAs belonging to different species, including miRBase (Griffiths-Jones et al., 2006; 2008), TarBase (Sethupathy et al., 2006; Vlachos et al., 2014), and miRTarbase (Chou et al., 2016). miRBase is an online database which is available at http://www.mirbase.org/ (Griffiths-Jones et al., 2006). The miRBase is an online searchable archive of published miRNA sequences and annotation. Each record in miRBase signifies a predicted hairpin-portion of a miRNA transcript called as 'mir' along with information of location and sequence of the mature miRNA. The miRBase also stores sequences of all the published mature miRNA, along with their predicted source hairpin precursors and annotation relating to their discovery, structure and function. miRBase has a nomenclature scheme for all predicted targets, for example, has-miR-121, in which the first three alphabets signify the organism, 'R' in miR denotes the mature miRNA sequence, and a number as a suffix. TarBase is a manually curated and experimentally supported collection of miRNA targets (Sethupathy et al., 2006). DIANA-TarBase v7.0 (Vlachos et al., 2014) stores more than half a million miRNA-gene interactions which uses 356 different cell types from 24 species, including human, mouse, fruit fly, zebrafish, and worms. miRTarBase is another manually curated database which stores more than 360 thousand miRNA-target interactions (Chou et al., 2016). These accumulated target-interactions have been further experimentally validated by reporter assay, western blot, microarray and next-generation sequencing experiments. miRTarBase release 6.0 (Chou et al., 2016) contains 3,786 miRNAs and 22,563 targets from 18 different species. There are several other databases such as miRDB (Wang, 2008), which uses miRNA sequences from miRBase and mRNA 3'-UTR sequences are imported from the GenBank files using BioPerl (http://www.bioperl.org), and uses MirTarget version 2 tool for the genome-wide target prediction (Wang and Naqa, 2007). miRNAMap (Hsu et al., 2006) is another database which stores the miRNA genes, putative miRNA genes, known and putative miRNA targets of human, mouse, rat and dog. The putative miRNA targets are obtained using RNAz (https://www.tbi.univie.ac.at/software/RNAz/), which is a tool used for non-coding RNA prediction based on comparative sequence analysis (Washietl et al., 2005). The mature miRNA of the putative miRNA genes are accurately predicted using a machine learning approach, called mmiRNA. The miRNA targets within the conserved regions of 3'-UTR of the genes are predicted using the miRanda algorithm (Enright et al., 2003). The miRGate (Andrés-León et al., 2015) is another comprehensive database consist of miRNA-mRNA pairs which are calculated using five target prediction algorithms: miRanda (Enright et al., 2003), TargetScan (Lewis et al., 2005; Bartel, 2009; Agarwal et al., 2015), RNAhybrid (Krüger & Rehmsmeier, 2006), microTar (Thadani& Tammi, 2006), and PITA (Kertesz et al., 2007). It also consists of complete sequences of miRNA and mRNAs 3'-UTRs of human (including human viruses), mouse, and rat with experimentally validated data. In miRGate, miRNA sequences are taken from the miRBase 20 (Kozomara and Griffiths-Jones, 2013) as it consists of a lot of datasets as compared to the other datasets. The miRGate obtained experimentally validated data from four databases: miRecords (Xiao et al., 2009), TarBase (Vergoulis et al., 2012), OncomirDB (Wang et al., 2014), and miRTarBase (Chou et al., 2016).

## 4. Materials and Methods

### 4.1 Datasets

For a comprehensive evaluation of predicted targets of miRNA from eleven different prediction tools, we have considered the miRNAs which were validated by experimental methods taken from miRNA databases to obtain optimal results. In this study, we have considered datasets from two species: *Drosophila melanogaster* and *human*. The high confidence miRNAs were downloaded from miRBase and their validated targets were obtained from the miRTarbase. The database consists of 28,645 entries of around 110 species at the time of writing this manuscript. The downloaded two datasets have been considered as a benchmark for the evaluation of eleven considered miRNA target prediction tools. Detailed description of these two datasets is as follows:

**Dataset-I: *Drosophilla melanogaster***

*Drosophilla melanogaster* [BDGP5.0] is a model organism, having total 256 precursor miRNA sequences. We have focused our study on the miRNAs having a high probability of expression level notated as high confidence miRNAs. *Drosophilla melanogaster* has 76 high confidence miRNAs. These high confidence miRNAs were then searched for their validated targets using miRBase and miRTarBase which is a database of experimentally validated microRNA-target interactions. According to miRTarBase, *Drosophilla melanogaster* shows 147 miRNA-target interactions between 45 miRNAs and 86 target genes. In this study, targets of *Drosophila melanogaster* miRNAs were predicted usingseven different tools,namely TargetScan, MicroT-CDS, PicTar, miRror, microRNA, ComiR, and PITA, and their performance evaluation has been performed.

**Dataset-II: *Human* (*Homo sapiens*)**

The name and sequences of highly confidence, mature miRNAs were downloaded from miRBase. There are 2588 highly confidence human miRNAs, out of which 208 random miRNAs were selected in this study. The validated target genes of all mature miRNAs were downloaded from TarBase and miRTarBase. These targets for human miRNAs were separated into another file using a program, which were used as the benchmark for testing ten target prediction tools, namely TargetScan, miRSystem, mirWalk, miRmap (Vejnar et al., 2012), miRSearch (Lewis et al., 2005; García et al., 2011), microT, microRNA, PITA, CoMir, and PicTar. The targets for each miRNA were predicted using considered ten tools and then further assessed for their performance and accuracy. Table 1 presents a brief summary of considered datasets for our comprehensive assessment.

**Table 1** Summary of datasets used for the assessment of miRNA target prediction tools

| Organisms | Number of miRNAs | Number of targets | Data source | Tools used for assessment |
|---|---|---|---|---|
| *Drosophila Melanogaster* | Out of 76 entries in miRBase, 44 experimentally validated are considered | 140 | miRBase, TarBase, MirTarBase | • PicTar (Krek et al., 2005)<br>• PITA (Kertesz et al., 2007)<br>• microRNA (Betel et al., 2008)<br>• CoMir (Coronnello&Benos, 2013)<br>• microT-CDS (Paraskevopoulou et al., 2013)<br>• MiRorSuite (Friedman et al., (2014)<br>• TargetScan (Agarwal et al., 2015) |
| *Human* | Out of 2,588 high confident miRNAs in miRBase, 208 are considered randomly | 26,315 | miRBase, TarBase, MirTarBase | • PITA (Kertesz et al., 2007)<br>• microRNA (Betel et al., 2008)<br>• miRSearch(Lewis et al., 2005; García et al., 2011)<br>• miRSystem (Lu et al. (2012)<br>• miRmap (Vejnar et al., (2012)<br>• microT-CDS (Paraskevopoulou et al., 2013)<br>• CoMir (Coronnello&Benos, 2013)<br>• mirWalk (Dweep et al., 2014)<br>• TargetScan (Agarwal et al., 2015)<br>• PicTar |

## 4.2 Target Prediction Tools

Categorizing the gene targets of miRNAs is essential for illustrating the biological mechanisms underlying these powerful regulatory molecules. There are several miRNA target prediction algorithms which exploit different approaches to predict the binding targets. In animal genomes, miRNAs show only partial complementarity to their target mRNA in disparity to plants where miRNAs can bind with almost perfect complementarity to their targets (Carrington and Ambros, 2003), which also makes it difficult to predict the target genes for animal genomes (Martin et al., 2007). In fact, these tools still need many fold of improvement and bioinformatics techniques require high-throughput experiments in order to validate predictions. Existing miRNA target prediction tools applies machine learning methods and probabilistic learning algorithms in order to construct predictive models whose foundation lies on experimentally verified miRNA targets. In the following section, we discussed miRNA target prediction techniques and summarized a detailed comparison of methodologies and features they use. All computer-based miRNA target prediction programs are created with specific features and parameters, where minor variation may result differently for the same input. The eleven prediction tools considered in this study are described briefly in the following section.

### 4.2.1 ComiR

ComiR (Combinatorial microRNA) is a web server to predict the targets of a set of miRNAs (Coronnello et al., 2012; Coronnello&Benos, 2013). It is easy to access and give the expecting result with higher accuracy in comparison to other tools. CoMir computes the potential of an mRNA being targeted by a miRNA in the species (human, mouse, fly, and worm genomes). The target genes can be predicted in two ways, either by entering a set of miRNAs along with their expression levels, or by entering a list of miRNA IDs. In the former case, CoMir calculates the targeting potential in two ways: first by applying four different methods: (i) miRanda (Enright et al., 2003) which calculates the probability of mRNA:miRNA binding based on the Fermi-Dirac equations (Zhao et al., 2009; Coronello et al., 2012) that consider the miRNA expression, and sum the individual probabilities over all of the mRNA of all miRNAs in the given set;  (ii) second method is similar to PITA (Kertesz et al., 2007), in which the equations substitutes the standard energies, (iii) in the third method, TargetScan (Lewis et al., 2005) scoring (without conservation) is weighted by each miRNA expression level, and (iv) mirSVR (Betel et al., 2010) is used, whose scores are combined to the weighted miRNA expression levels. Finally, in the second step, the predictions of the above four methods applied in the first step are combined with the support vector machine (SVM) which is trained on high quality dataset. On the other hand, when the miRNA IDs are input without expression levels, the CoMir assumes all the miRNAs as expressed at the same level (Coronello and Benos, 2013). If single miRNA is selected, ComiR computes each gene targeting score for miRNA for selected species. It has an optional box in which we can input single miRNA sequence in FASTA format and it will predict all target genes for the miRNA. All required mature miRNA sequences can be downloaded from miRbase database. ComiR supports four species: *H. sapiens, D. melanogaster, E. elegans and M. musculus.* Evaluation of the result can be done in two manners either based on rank or score.

### 4.2.2 microT-CDS

DIANA-microT-CDS (Paraskevopoulou et al., 2013) is the latest version of microT algorithm. The algorithm uses the presence of a positive and negative set of miRNA recognition elements (MREs) to be found in both the 3'-UTR and CDS regions. DIANA-microT-CDS achieves a major rise in sensitivity as compared to previous versions. The sensitivity according to the available review literature of microT-CDS is ~65%, whereas it was only 52% in the older versions of microT. In our study, miRNA target prediction is made using microT-CDS on *Human* and *D. melanogaster* (fruit fly) high confidence miRNAs datasets.

### 4.2.3 MicroRNA

MicroRNA (Betel et al., 2008) is an online target prediction tool, which predicts candidate targets and its downregulation scores using the mirSVR (Betel et al., 2010) machine learning method. The miRanda algorithm (Enright et al., 2003) is used to predict the targets and observed miRNA expression level. It computes the complementarity between a given set of miRNAs and an mRNA on the basis of weighted Smith-Waterman algorithm. The secondary filter applied in this tool is to estimate the free energy of the formation of the miRNA:mRNA duplex. The current version is used to predict targets for Human, *Drosophila melanogaster*, roundworm and mouse. Targets are predicted through miRNA identifiers and species. It gives all target genes with their alignment sites.

### 4.2.4 miRror Suite

miRror Suite (Friedman et al., 2010; 2014) is an online tool to predict likely targets for a set of miRNAs. It has two protocols for prediction: gene to miRNA and miRNA to gene. miRror ranks a list of target genes according to their likelihood to be targeted by the given set of miRNAs. It requires miRNA ID and gene accession ID to predict the expected results. It accepts a set of miRNAs/genes or at least two valid miRNA/genes. miRror supports several species and integrates many other resources, including TargetScan database (Grimson et al., 2007), PITA (Kertesz et al., 2007), PicTar (Krek et al., 2005), Microcosm (John et al., 2004), MiRanda (Betel et al., 2008) (conserved and non-conserved), miRDB (Wang, 2008), RNA22 (Miranda et al., 2006) and Mirz (Hausser et al., 2009). It gives scores based on the integrated databases.

### *4.2.5 PicTar*

PicTar is an algorithm for the identification of miRNA targets (Krek et al., 2005). It supports several organisms including *Drosophila*, and the non-conserved co-expressed human miRNAs.After entering a query (nucleotide sequence of mature miRNA or multiple sequence alignment of RNA residues), PicTar first locate all the possible sites termed as nuclei (length 7, starting at position 1 or 2 of the 5' end of the miRNA) in the given sequence followed by some filters. The optimal free energy of each nuclei is predicted which narrows down to lesser targets. The highly probable nuclei with optimal free energy falling into the overlapping positions in the alignment of the considered species are called anchors. If the 3'-UTR alignment has enough anchors, each UTR in the alignment is then subjected to be scored by the central PicTar maximum likelihood procedure, after which all the scores of the orthologous transcripts are combined. Finally, a list of transcripts ranked by the PicTar score is displayed (Krek et al., 2005).

### *4.2.6 PITA*

PITA (Kertesz et al., 2007) incorporates a new approach the prediction of miRNA targets. Its main hypothesis is based on the fact that mRNA structure plays significant role in recognizing targets by thermodynamically promoting or suppressing the interaction. This tool allows the user to run the PITA algorithm on his choice of UTRs and miRNAs. PITA first scans the UTR for potential miRNA targets and then scores each site according to the parameter-free model explained by Kertesz et al., (2007). This model computes the difference between the free- energy gained by the formation of miRNA-target duplex and the energy released by the un-pairing of the target to make it accessible to the miRNA. The PITA algorithm uses the features such as seed match, free energy, site accessibility, target-site abundance, and G:U pairs allowed in the seed.

### *4.2.7 TargetScan*

TargetScan (Lewis et al., 2005; Bartel, 2009; Agarwal et al., 2015) is one of the wide-range miRNA target prediction tool that supports human, mouse, fruit-fly, worm, and fish. It has been upgraded several times and provides wide range of information about their predicted as well as validated binding sites on their target genes. It estimates the cumulative weighted context++ score (CWCS) for each miRNA. The CWCS score ranks based upon the predicted repression or PCT (probability of conserved targeting) aggregated score of the longest 3'-UTR isoform. Firstly, the 6mer, 7mer-A1, 7mer-m8, and 8mer are first filtered to remove overlapping sites for each miRNA family, then the CWCS is calculated for each member of a miRNA family, and the member which represents the greatest predicted repression score, is chosen to represent that family and the reference 3'-UTR with the most 3p-seq tags represents the gene (Agarwal et al., 2015).

### *4.2.8 miRSystem*

miRSystem (Lu et al., 2012) integrates seven tools to predict targets for miRNAs, which includes DIANA-microT (Maragkakis et al., 2009), miRanda (Betel et al., 2008), miRBridge (Tsang et al., 2010), PicTar (Krek et al., 2005), PITA (Kertesz et al., 2007), RNA22 (Miranda et al., 2006), and TargetScan (Lewis et al., 2005; Bartel, 2009; Agarwal et al., 2015). Currently it supports human and mouse.

### *4.2.9 miRWalk*

miRWalk (Dweep et al., 2014) is a comprehensive miRNA target prediction tool, which integrates 12 existing target prediction tools, namely DIANA-microTv4.0, DIANA-microT-CDS, miRanda Release 2010, mirBridge (Tsang et al., 2010), miRDB4.0 (Wang, 2008), miRmap (Vejnar et al., 2012), miRNAMap (Hsu et al., 2006), PicTar2, PITA, RNA22 version 2, RNAhybrid2.1 (Krüger & Rehmsmeier, 2006), and TargetScan6.2. It provides the miRNA binding sites within the complete sequence of a gene. It supports human, rat, dog, mouse, and cow species.

### *4.2.10 miRMap*

miRMap (Vejnar and Zdobnov, 2012) is a comprehensive prediction tool which implements eleven different features for target prediction. One of the eleven featuresevaluates the significance of negative selection, which is based on a performing predictor for evolution named PhyloP (Pollard et al., 2010). Currently, it supports human, mouse, rat, cow, opossum, chicken, chimpanzee, and zebrafish.

### *4.2.11 miRSearch*

miRSearch (Lewis et al., 2005; García et al., 2011) is an online search tool for miRNA targets interaction. The results are based on TargetScan (Lewis et al., 2005) providing gene targets for human, mouse, and rat miRNAs. miRSearch uses an advanced algorithm to cross-reference all the annotations found in the literature and displays a comprehensive list of miRNA-mRNA interactions. It uses context++ score to predict results without considering site conservation.

All the 11 considered target prediction tools are compared in terms of input requirement, tool features, supported species, tool URL, and its citation, as shown in Table 2. The objective of this paper is to evaluate and assess the performance of miRNA target prediction tools in human and *drosophila melanogaster*. Therefore, we have considered those tools which either supports *human* or *drosophila melanogaster*.

**Table 2. List of tools for miRNA target prediction**

| S.No. | Name of Tools | Input | Tool features | Supported species | Tool URL | References |
|---|---|---|---|---|---|---|
| **1.** | ComiR | miRNA name | <ul><li>seed match</li><li>conservation</li><li>free energy</li><li>site accessibility</li><li>target-site abundance</li><li>machine learning</li><li>3' compensatory pairing</li><li>G:U pairs allowed in the seed</li><li>local AU content</li><li>miRNA expression level</li></ul> | <ul><li>Human</li><li>Mouse</li><li>Fly</li><li>Worm</li></ul> | http://www.benoslab.pitt.edu/comir/ | Coronnello et al., (2012). |
| **2.** | microT-CDS | miRNA name, gene name, Ensembl ID | <ul><li>seed match</li><li>conservation</li><li>free energy</li><li>site accessibility</li><li>target-site abundance</li><li>machine learning</li><li>3' compensatory pairing</li><li>G:U pairs allowed in the seed</li><li>local AU content</li></ul> | <ul><li>Human</li><li>Mouse</li><li>Fly</li><li>Worm</li></ul> | http://diana.imis.athena-innovation.gr/DianaTools/index.php?r=microT_CDS/index | Paraskevopoulou et al., (2013). |
| **3.** | microRNA | miRNA name | <ul><li>local AU content</li><li>seed match</li><li>conservation</li><li>secondary structure accessibility</li></ul> | <ul><li>Human</li><li>Mouse</li><li>Fly</li><li>Rat</li></ul> | http://www.microrna.org/microrna/home.do | Betel et al. (2008). |
| **4.** | MiRorSuite | miRNA name | <ul><li>seed match</li><li>conservation</li><li>free energy</li><li>site accessibility</li><li>target-site abundance</li><li>machine learning</li><li>3' compensatory pairing</li><li>G:U pairs allowed in the seed</li><li>local AU content</li></ul> | <ul><li>Human</li><li>Mouse</li><li>Fly</li><li>Rat</li><li>Worm</li><li>Fish</li></ul> | http://www.protocs.huji.ac.il/mirror/index.php | Friedman et al., (2014). |
| **5.** | Pictar | miRNA name, gene name | seed match, pairing stability | <ul><li>Vertebrate</li><li>Fly</li><li>Worm</li><li>Mouse</li></ul> | http://pictar.mdc-berlin.de/ | Krek et al., 2005 |
| **6.** | PITA | miRNA name | <ul><li>seed match</li><li>free energy</li><li>site accessibility</li><li>target-site abundance</li><li>G:U pairs allowed in the seed</li></ul> | <ul><li>Human</li><li>Mouse</li><li>Fly</li><li>Worm</li></ul> | https://genie.weizmann.ac.il/pubs/mir07/index.html | Kertesz et al., (2007). |

| 7. | TargetScan | miRNA name, miRNA family, gene name | • seed match<br>• conservation<br>• free energy<br>• site accessibility<br>• target-site abundance<br>• 3' compensatory pairing<br>• G:U pairs allowed in the seed<br>• local AU content | • Human<br>• Mouse<br>• Fly<br>• Worm<br>• Fish | http://www.targetscan.org/vert_71/ | Lewis et al., 2005; Bartel, 2009; Agarwal et al., 2015 |
| 8. | miRmap | miRNA name | • seed match<br>• conservation<br>• free energy<br>• site accessibility<br>• local AU content<br>• Site over-representation probability | • Human<br>• Mouse<br>• Rat<br>• Cow<br>• Chicken<br>• Zebrafish | http://mirmap.ezlab.org/ | Vejnar et al., (2012) |
| 9. | miRSearch | miRNA name | miRSearch uses an advanced cross-referencing system to identify validated and predicted miRs for any target. | • Human<br>• Mouse<br>• Rat | https://www.exiqon.com/miRSearch | Lewis et al. (2005); García et al. (2011) |
| 10. | miRSystem | miRNA name | Integrated system using seven prediction tools: DIANA, miRanda, miRBridge, PicTar, PITA, rna22, and TargetScan. | • Human<br>• Mouse | http://mirsystem.cgm.ntu.edu.tw/ | Lu et al. (2012). |
| 11. | miRWalk | miRNA name | Combines 12 existing miRNA target prediction algorithms: DIANA-microTv4.0, DIANA-microT-CDS, miRanda-rel2010, mirBridge, miRDB4.0, miRmap, miRNAMap, doRiNA i.e., PicTar2, PITA, RNA22v2, RNAhybrid2.1 and Targetscan6.2 | • Human<br>• Mouse<br>• Rat | http://129.206.7.150/ | Dweep et al., 2014 |

## 4.3 Empirical evaluation

We used comprehensive evaluation metrics to analyze the performance and accuracy of the considered eleven miRNA target prediction tools. The predicted targets were categorized into four categories: true positives (TP), false positives (FP), true negatives (TN), and false negatives (FN). TP and TN refers to the count of the correctly predicted functional and non-functional targets, respectively, whereas, FP and FN are the counts of the functional and non-functional targets which were not validated by the experimentally proven targets. The predictions were assessed using the following measures:

$$Precision = \frac{TP}{TP + FP} \qquad (1)$$

$$Recall/TPR = \frac{TP}{TP + FP} \qquad (2)$$

$$FPR = \frac{FP}{FP + TN} \qquad (3)$$

$$Average\ predictions\ per\ miRNA = \frac{total\ no.\ of\ targets\ predicted}{total\ no.\ of\ miRNAs\ considered} \qquad (4)$$

$$F - measure = 2 * \frac{Precision * Recall}{Precision + Recall} \qquad (5)$$

## 5. Results and Discussions

We evaluated the performance of eleven different miRNA target predictors on the datasets of two different species (human and *D. melanogaster*), scaling them on different parameters such as sensitivity, specificity, precision, recall, and other similar empirical methods. The strategy applied to evaluate the performance of miRNA target predictors is shown in Fig. 1. The performance evaluation of prediction tools was performed in terms of different metrics namely, average prediction per miRNA, TPR, F-measure, and combining results of best performing tools (union & intersection) on experimentally validated *Drosophila Melanogaster* and Human miRNA targets.

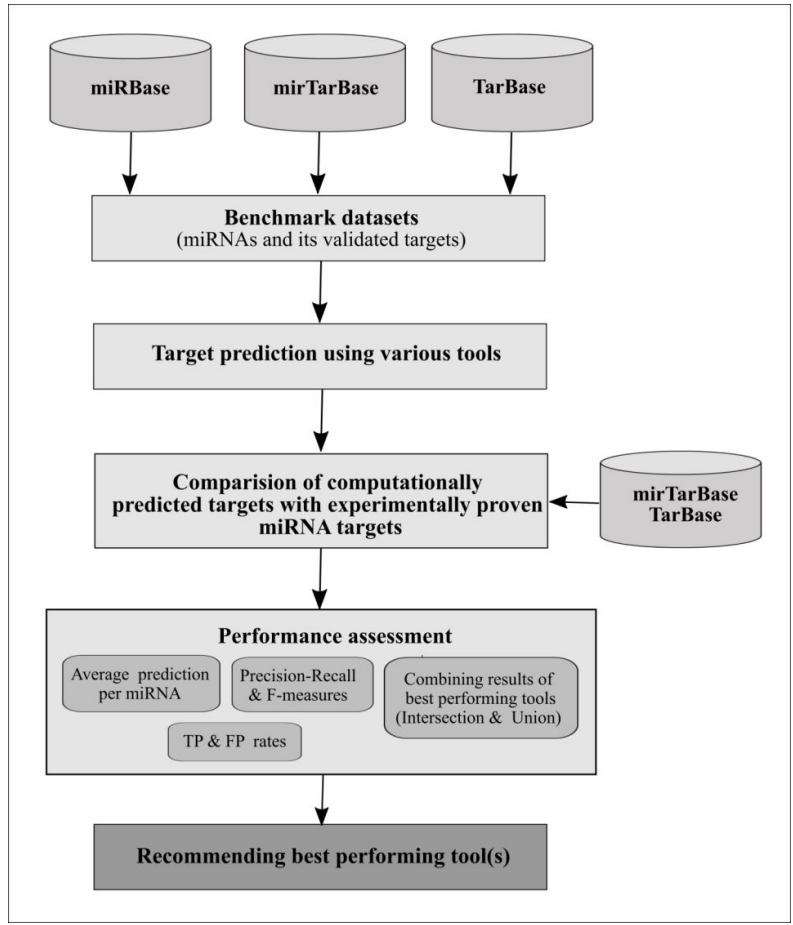

**Fig. 1** Representation of the strategy applied to evaluate mRNA target prediction tools

## 5.1 Dataset-I (*Drosophila*)

### (a) *Average predictions per miRNA*

A single miRNA could target multiple (6-7) mRNAs and a single mRNA could be targeted by several (4-5) miRNAs (German et al., 2008; Beauclair et al., 2010). Our results shown in Fig. 2 are consistent with this hypothesis. The average number of targets predicted for *D. melanogaster* ranged between 100-1400 per miRNA (Fig. 2). The results suggest that the tool microRNA predicted the highest number of targets which is followed by CoMir, microT, miRor, PITA, PicTar, and TargetScan.

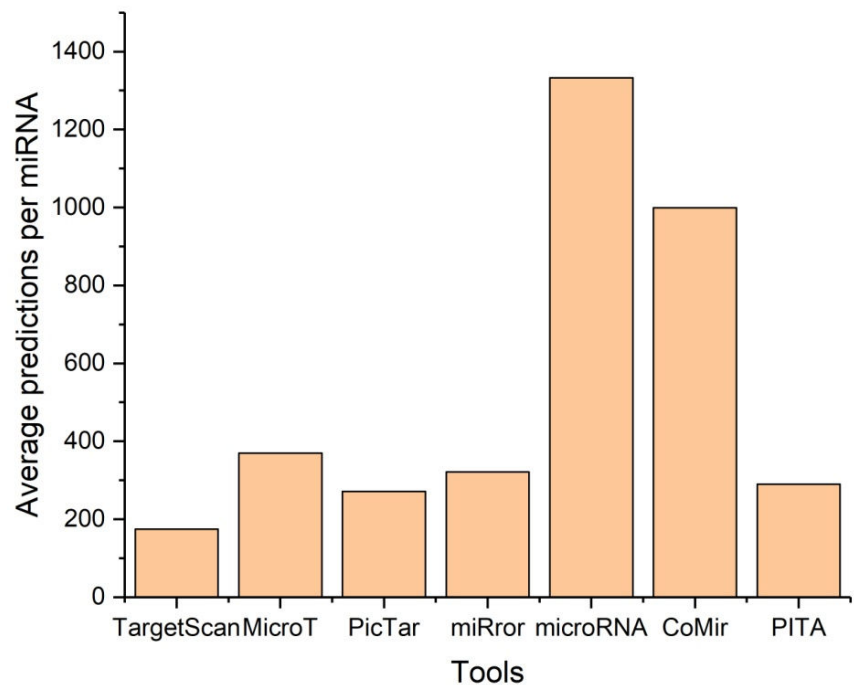

**Fig. 2** Average number of target prediction per miRNA by different tools in *Drosophila.*

### (b) *True positive rate (TPR) & false positive rate (FPR)*

The plot of TPR and total number of predicted targets are shown in Fig. 3, which suggests that the seven considered tools for *drosophila* dataset predicted a large number of targets and majority of the tools followed the similar pattern. The highest TPR is achieved by the tool MicroT followed by Comir with a very slight difference. All the considered tools predict a large number of false positives, and due to that the FPR goes to ~99%.

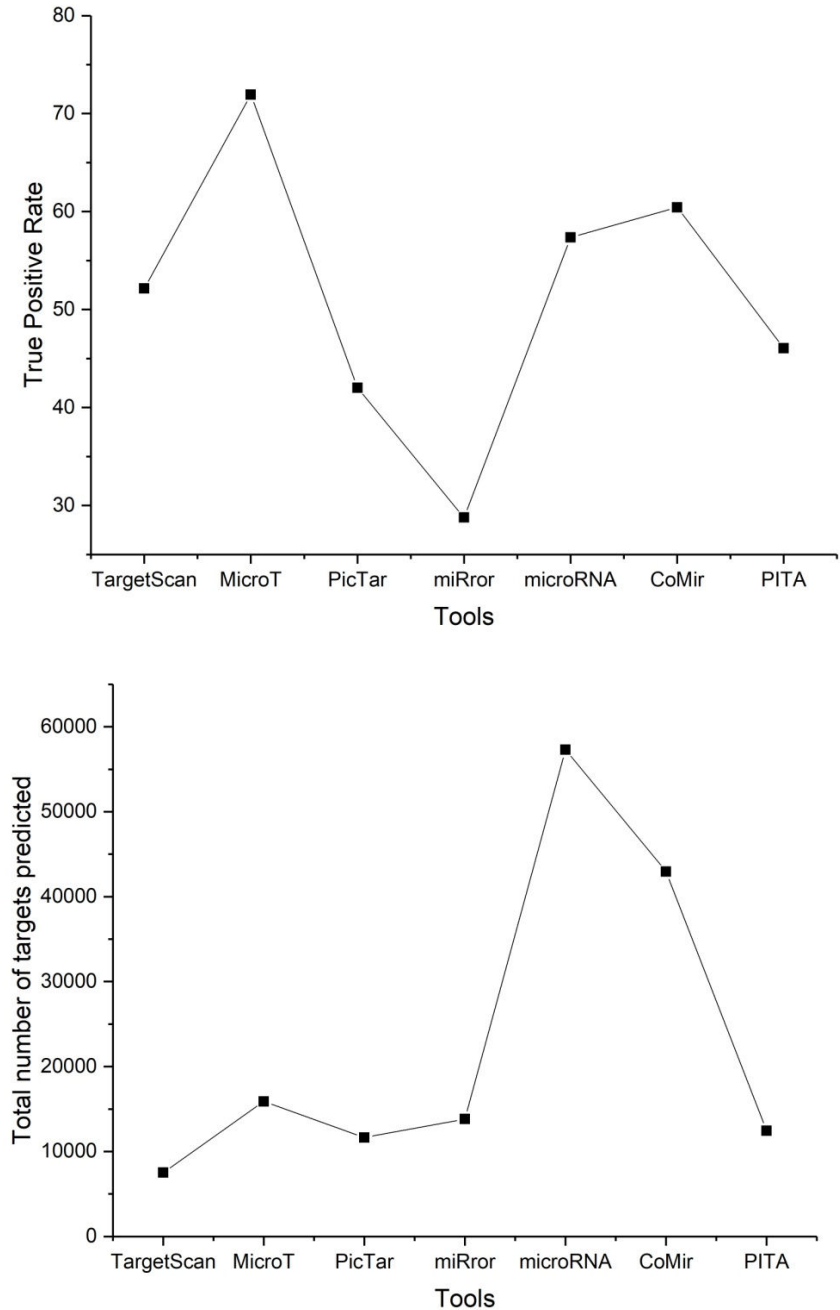

**Fig. 3** Evaluation of tools on the basis of true positive rate (TPR) and total number of targets predicted for *Drosophila melanogaster.*

## *(c) Precision-recall and F-measure*

Precision and recall are the important parameters to evaluate the accuracy and sensitivity of predictions.. According to the results obtained from the analysis of *drosophila* dataset, TargetScan showed the highest precision 0.0097 (though smaller according to the scale) and the recall is 0.5214 (Supplementary file). The precision of these seven tools ranged between 0.005 to 0.009 and recall ranges between 0.2 and 0. The highest recall is shown by the MicroT tool whereas the precision comes next to that of TargetScan (Fig. 4). The tools were evaluated at an optimal score of 0.0 and the F-measure was calculated which is a harmonic mean of precision and recall, also known as the F1-score. The F-measure is only high when the precision and recall are high. The highest F-measure amongst the seven tools used to predict targets for *Drosophila melanogaster* is shown by TargetScan followed by MicroT.

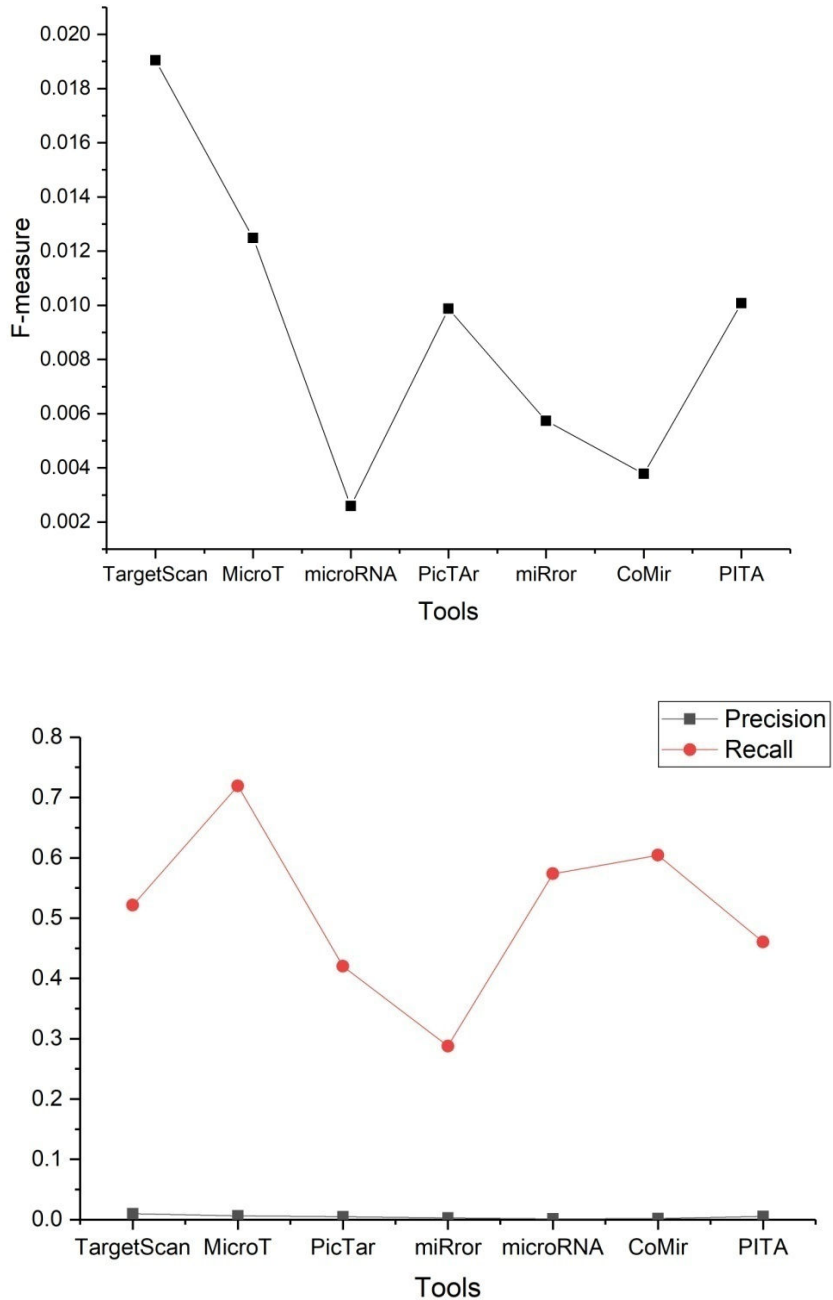

**Fig. 4** A line graph showing the F-measure and Precision-recall calculated *drosophila* dataset. TargetScan showed the highest F-measure, while microT shows the highest recall.

### (d) Combining results to improve accuracy and reliability

It is hypothesized that unions of predicted results are supposed to achieve higher recalls when compared to the outcomes of individual tools. Similarly, the intersections may achieve higher precisions. The results of the best performing tools for *drosophila* dataset were combined as unions and intersections to improve their recall and precision (Fig. 5). The union of MicroT and microRNA showed a two fold increase in the TPs, which was as much as found to be decreased in the case of the intersection of microRNA and CoMir.

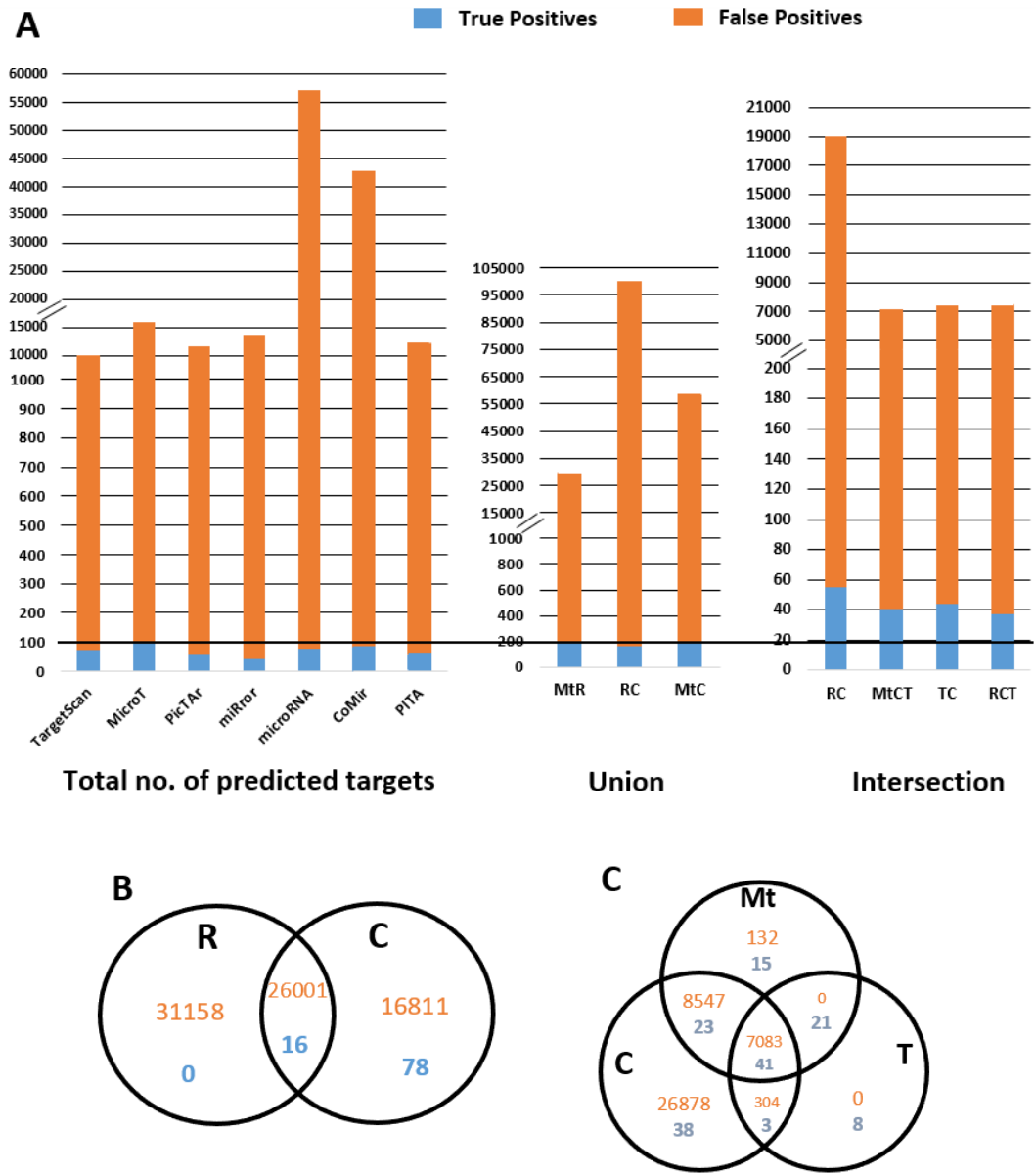

**Fig. 5** Combined outputs of the dataset-I (*D. melanogaster*) to improve accuracy & reliability. **A)** represents the comparison of the total miRNA targets predicted by the tools to the union and intersection of the tools to improve their performance. **B)** union of microRNA and Comir, and **C)** intersection of MicroT, Comir and TargetScan. Abbreviations: Mt–MicroT, R–microRNA, C–Comir, T–TargetScan.

## Dataset-II (*human*)

### (a) *Average predictions per miRNA*

In the case of *human* dataset, the average number of predicted targets ranged between 100-8000 per miRNA (Fig. 6). The results suggests that miRmap predicted the highest number of targets which is followed by TargetScan, Comir, miRWalk, MicroT, PITA, miRSearch, microRNA, and miRSystem.

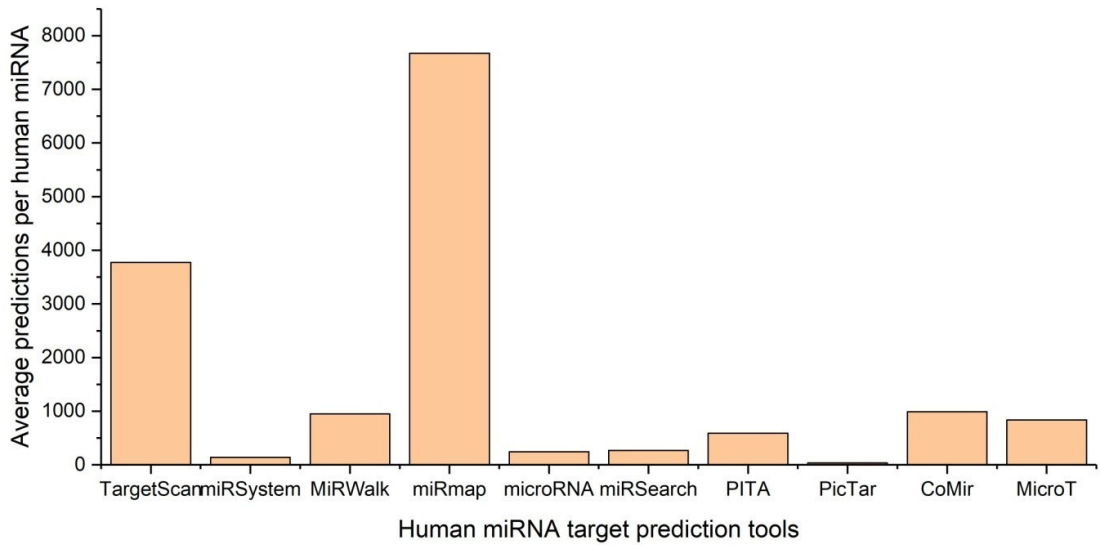

**Fig. 6** Average number of target prediction per miRNA by different tools in the *human* dataset

### (b) *True positive rate (TPR)*

In the case of *human* dataset, the plot of TPR, FPR, and the total number of predicted targets for *human* dataset is shown in Fig. 7, which suggests that the seven tools predicted a large number of targets. According to the results, the highest TPR was achieved by the TargetScan followed by miRMap with a small difference. However, miRMap predicted higher number of targets than the TargetScan (Fig. 7A). This shows that the variation between the TPR and the total number of predicted targets (Fig. 7B) is inconsistent and independent. It means that whether a tool predicts higher or smaller number of targets, the TPR remains unaffected. The least FPR is achieved by the tool CoMir followed by MicroT, and TargetScan, miRSystem, miRWalk, miRmap, microRNA, miRSearch, and PITA attained the FPR closer to each other followed by PITA. According to the results, the TargetScan showed the higher TPR and FPR as well.

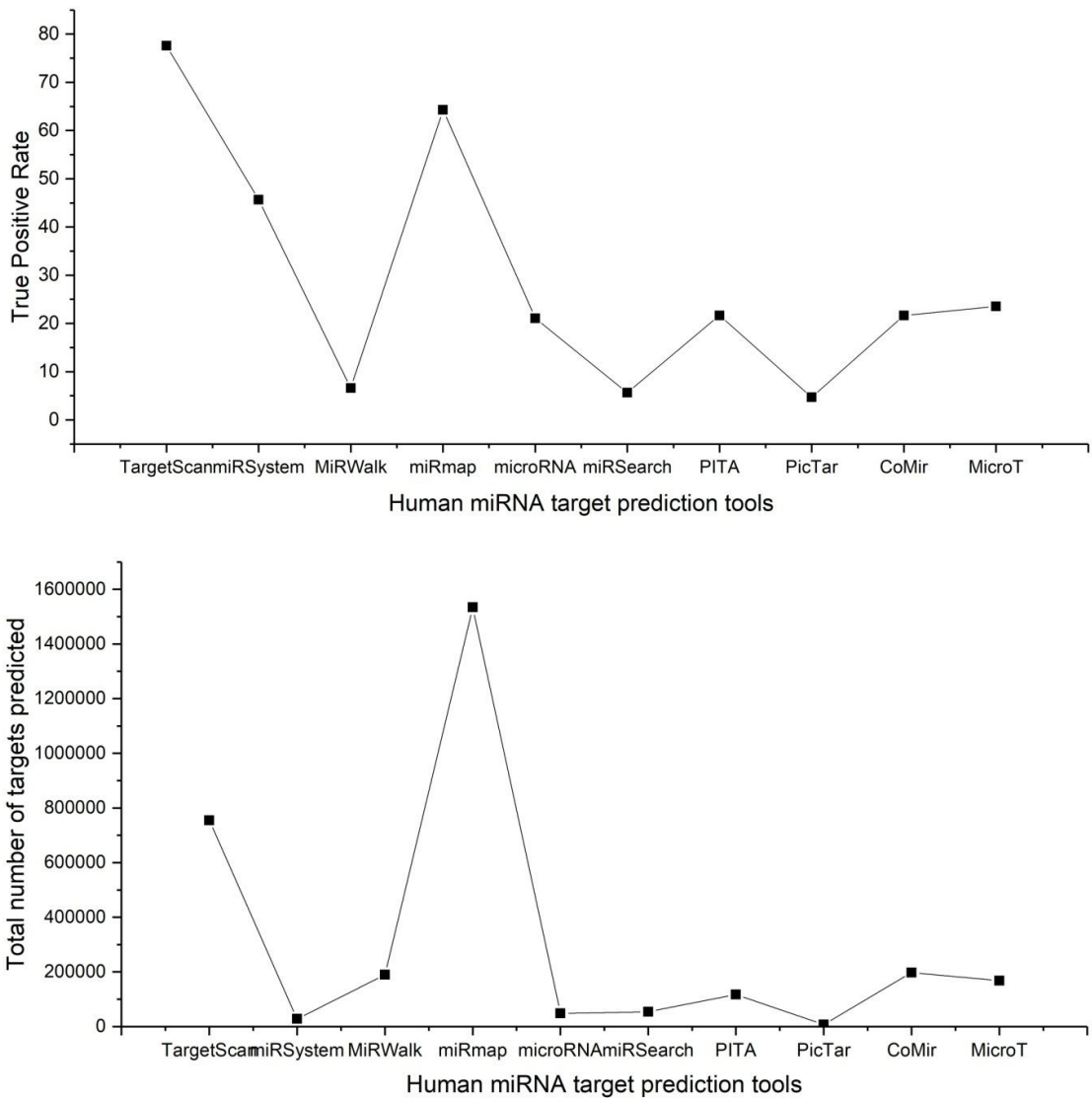

**Fig. 7** TPRs and total number of targets predicted by different tools

## (c) Precision-recall and F-measure

The precision and recall of *human* dataset was also calculated. According to the results, miRSearch showed the highest precision 0.03 (though smaller according to the scale) while the highest recall was showed by the TargetScan (0.77) (Supplementary file). The precision and recall of these seven tools ranged between 0.009 to 0.03 and 0.05 to 0.77 respectively. Since the range of the precision and recall was very low, therefore, we calculated the F-measure for the tools (Fig. 8). The highest F-score was shown by MicroT followed by the TargetScan, CoMir, miRSystem, miRSearch, PITA, PicTar, miRmap, and miRWalk.

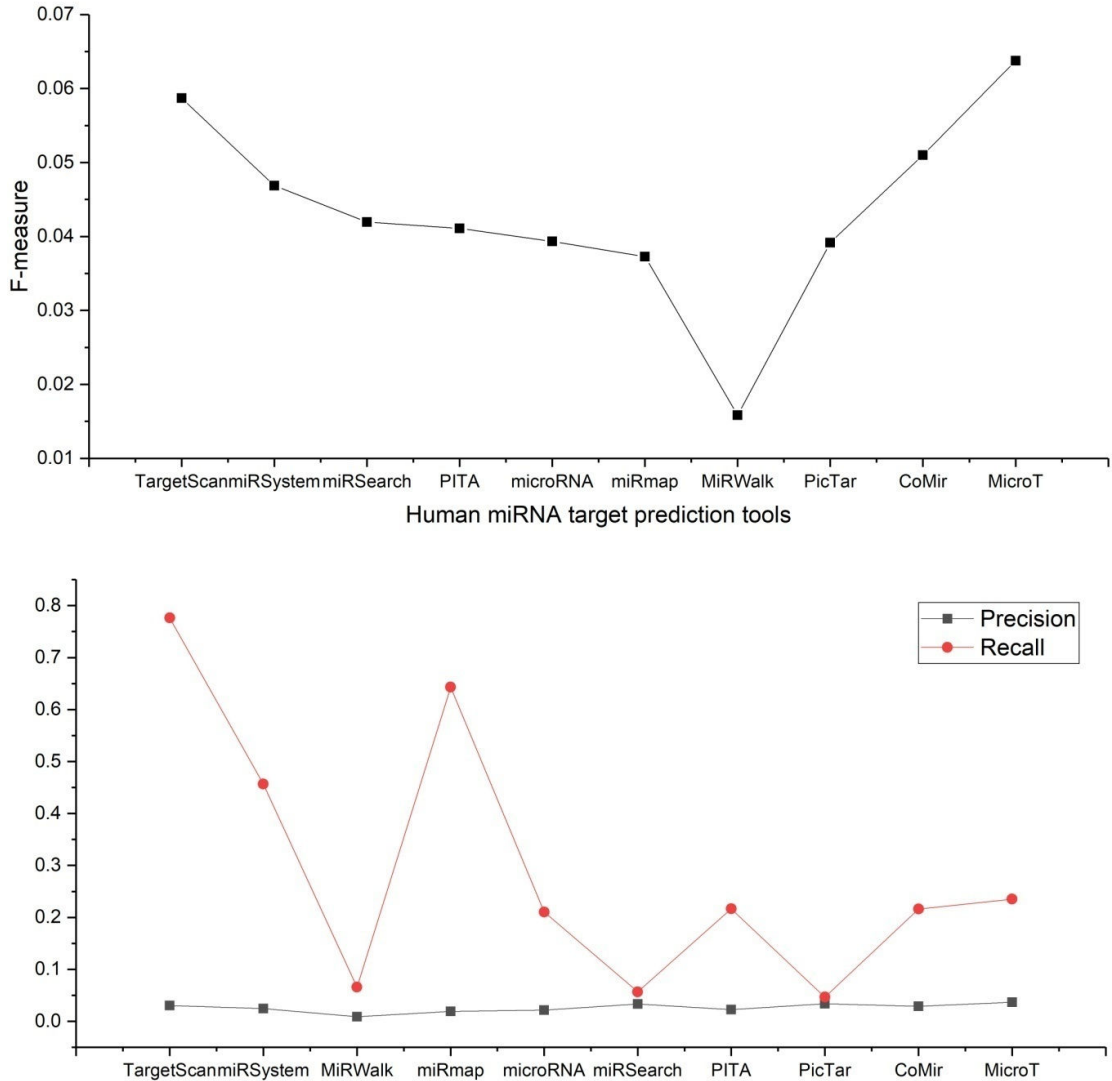

**Fig 8** F-measure and Precision-recall of human dataset

### (d) Combining results to improve accuracy and reliability

The results of the best performing tools considered for *human* dataset were combined as unions and intersections to increase their performance and accuracy, as done for the first dataset of *D. melanogaster*. After score optimizations, the combination of TargetScan and miRmap resulted in 31914 TPs, which is a higher than the TPs predicted by the TargetScan alone (Fig. 9). The combination of miRmap and Comir was less than that of the former, but the false positives were more than that of the TargetScan. The intersection of miRmap, Comir, and miRSearch were very low, which showed only 162 TPs shared by these three tools.

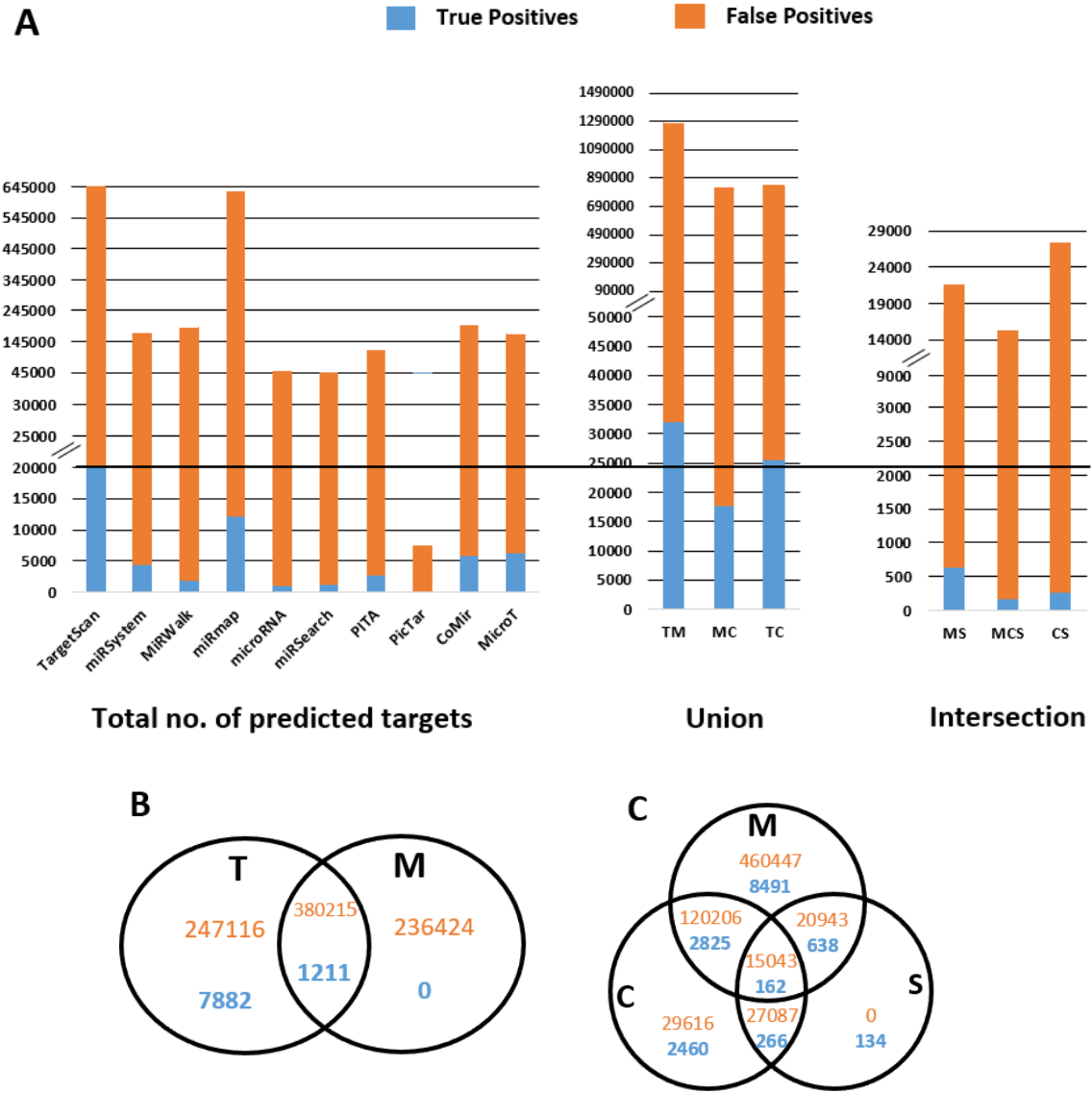

**Fig. 9** Combined results on dataset-II (*human*) to improve accuracy. TargetScan showed the highest number of the TPs and the FPs. The union of PITA and miRMap is lesser than the predictions of TargetScan alone. The intersection of MCS showed very less number of predicted targets. Abbreviations: M–miRmap, R–microRNA, C–Comir, T–TargetScan.

# 6. Conclusion

We analyzed eleven miRNA target predictors on two benchmark datasets by applying significant empirical methods to evaluate and assess their accuracy and performance. The best performing tools for the datasets evaluated on the basis of metrics shown in Table 3. According to our results, MicroT, microRNA, and CoMir showed the highest performance in dataset-I (*Drosophila melanogaster*), and in the dataset-II (*Human*), TargetScan and miRMap showed the best performance. The predicted results were combined to improve the performance of the tools in both the datasets, but any relevant improvement was not observed in the TPs. It was also observed that the TPR is independent of the number of targets predicted by a tool. For example, in the case of dataset-II (*human*), miRWalk predicted a large number of targets, but the TPR was very low.

**Table 3** Metric evaluated best performing tools for both the datasets.

| Evaluation methods | Datasets | |
|---|---|---|
| | **Dataset-I (*D. melanogaster*)** | **Dataset-II (*human*)** |
| True positive rate | MicroT | TargetScan |
| F-measure | TargetScan | MicroT |
| Precision-Recall | MicroT | TargetScan |
| Union of results | MicroT-microRNA | TargetScan–miRMap |

On the basis of the previous and our analysis, we can say that the existing tools have many limitations and drawbacks which embark the need for more accurate and precise miRNA target predictors. The current tools generate a large amount of false positives and works on different algorithms which makes it difficult to compare them. Although several algorithms and models have been developed to predict miRNAs *in-silico,* prediction of significant targets with high statistical confidence is still a challenging task.

**Conflict of Interest**

The authors declare that there is no any conflict of interest in the publication of the manuscript.

**Acknowledgement**